# Mechanical resonant sensing of spin texture dynamics in a two-dimensional antiferromagnet

S M Enamul Hoque Yousuf[1], Yunong Wang[1], Shreyas Ramachandran[2], John Koptur-Palenchar[3], Chiara Tarantini[4], Li Xiang[4], Stephen McGill[4], Dmitry Smirnov[4], Elton J. G. Santos[2,5,6]\*, Philip X.-L. Feng[1,3]\*, Xiao-Xiao Zhang[3]\*

[1]*Department of Electrical & Computer Engineering,
University of Florida, Gainesville, Florida 32611, USA*
[2]*Institute for Condensed Matter Physics and Complex Systems, School of Physics and Astronomy, The University of Edinburgh, Edinburgh EH9 3FD, UK*
[3]*Department of Physics, University of Florida, Gainesville, Florida, 32611, USA*
[4]*National High Magnetic Field Laboratory, Tallahassee, 32312, Florida, USA*
[5]*Donostia International Physics Centre (DIPC), Donostia-San Sebastian, 20018, Spain.*
[6]*Higgs Centre for Theoretical Physics, The University of Edinburgh, EH9 3FD, UK*

\**Emails: [xxzhang@ufl.edu](xxzhang@ufl.edu), [philip.feng@ufl.edu](philip.feng@ufl.edu), [esantos@ed.ac.uk](esantos@ed.ac.uk)*

## Abstract

The coupling between the spin degrees of freedom and macroscopic mechanical motions, including striction[1], shearing[2], and rotation[3-4], has attracted wide interest with applications in actuation, transduction, and information processing[5-7]. Experiments so far have established the mechanical responses to the long-range ordered[1, 5] or isolated single spin states[8]. However, it remains elusive whether mechanical motions can couple to a different type of magnetic structure, the non-collinear spin textures, which exhibit nanoscale spatial variations of spin (domain walls, skyrmions, *etc*.) and are promising candidates to realize high-speed computing devices[9-12]. Here, we report the detection of collective spin texture dynamics with nanoelectromechanical resonators made of two-dimensional antiferromagnetic (AFM) MnPS$_3$ with $10^{-9}$ strain sensitivity. By examining radio frequency mechanical oscillations under magnetic fields, new magnetic transitions were identified with sharp dips in resonant frequency. They are attributed to the collective AFM domain wall motions as supported by the analytical modeling of magnetostriction and large-scale spin-dynamics simulations. Additionally, an abnormally large modulation in the mechanical nonlinearity at the transition field infers a fluid-like response due to the ultrafast domain motion. Our work establishes a strong coupling between spin texture and mechanical dynamics, laying the foundation for electromechanical manipulation of spin texture and developing quantum hybrid devices.

# 1. Introduction

In most conventionally studied magnetic materials, the spin states establish long-range orderings and, therefore, can be simplified to a single or several spin sublattices, enabling them to be easily addressed by micro- and macroscopic approaches. In contrast, spin textures display nanoscale spatial variations of complex spin orientations and host a variety of exotic magnetic structures, including domain walls (DWs)[13], skyrmions[11], helical magnets[14], and spin liquids[15]. Spin texture excitations in antiferromagnets (AFMs), where the spin states have compensated magnetic moments, are particularly interesting and offer new perspectives in designing information processing and logic devices[9-12, 16-17]. The strong magnetic exchange interactions in AFM can lead to rich spin configurations and fast dynamics in the GHz to THz range[18]. However, detecting and manipulating these complex nanoscale spin textures and dynamics is particularly challenging due to the vanishing overall magnetic moments.

Recent advances in van der Waals (vdW) magnetic crystals offer new opportunities to investigate and manipulate spin interactions in low-dimensional systems[19-22]. Exotic spin textures can be found and on-demand engineered in various two-dimensional (2D) magnets and magnetic heterostructures like moiré structures[23-26] with potential electrical tunability[21]. In particular, the robust mechanical stabilities and outstanding elastic properties of the 2D vdW materials provide unique advantages for designing efficient nanomechanical devices and enable new device geometries that cannot be realized in 3D or conventional thin-film magnetic systems. Here, we report the mechanical sensing of nanoscale spin texture formation in 2D AFM materials enabled by exchange magnetostriction. Magnetostriction, which is the lattice change due to spin interactions, has applications in both classical[5] and quantum transduction and sensing[6-7]. Magnetometry based on such magneto-mechanical effects has been applied in single-spin and long-range ordered (ferromagnetic and AFM materials) magnetic systems. Examples include the sensing of a single electron spin[8, 27], ferromagnetic stochastic processes[28], and magnetic phase transitions[29-30] with magnetized mechanical cantilevers and membranes. In the formation of spin textures, the rearrangement of neighboring spins leads to changes in magnetic exchange interactions, especially in AFM materials with large exchange energies. The corresponding exchange-induced attraction or repulsion of neighboring sites thus yield a new equilibrium lattice position and macroscopic mechanical changes to minimize the system's total energy. While AFM spin textures and domains are conventionally difficult to detect due to the vanishing magnetization and nanoscale sizes, we expect high-sensitivity magnetostriction measurements to provide access to detecting the formation and dynamics of spin textures.

# 2. Results and Discussion

## 2.1 Mechanical Resonator Device Characterizations

To achieve ultrasensitive mechanical sensing of the spin texture, we fabricate 2D AFM material $MnPS_3$ into resonant nanoelectromechanical systems (NEMS) operating in the radio frequency (RF) range. $MnPS_3$ is a 2D Heisenberg-type van der Waals AFM below its Neel temperature of 78 K[31-34]. **Figure 1a** illustrates the corresponding spin structures. At zero magnetic field, the spins are fully compensated within each of the layers and collinearly aligned with a tilting angle of ~8° relative to the perpendicular direction[34]. Under an out-of-plane magnetic field, a gradual spin-flop (SF) transition is expected near 4-5 T. The magnetization of bulk MnPS3 was measured and included in Supplementary Section S2. The $MnPS_3$ NEMS resonators with low dissipation (therefore, a high quality factor $Q$) enable a high-sensitivity detection of magnetostriction effects under a varying field, as depicted in Figure 1b. The magnetic field dependence of the linear and nonlinear dynamics of the 2D AFM mechanical resonators exhibits two sharp magnetic transitions, one close to the expected SF field, and the other being a completely new transition at a higher field ~7 T. Analytical modeling and atomistic spin-dynamic simulations indicate that these sharp transitions arise from fast domain wall (DW) motions at the switching fields. The large modulation in macroscopic mechanical properties is attributed to the coupling of rapid local spin texture dynamics and mechanical resonant motions. Furthermore, the abnormal sign change of third-order (Duffing) and enhancement of fifth-order (quintic) mechanical nonlinearity terms suggest a fluid-like DW motion and dynamic process near the SF field. This work establishes a new detection scheme of nanoscale spin textures and opens up a pathway to manipulating novel spin excitations with engineerable mechanical resonance modes.

Mechanically exfoliated $MnPS_3$ flakes were transferred and suspended over circular cavities with varying diameters prefabricated on a sapphire substrate to form the NEMS structures with controlled mechanical degrees of freedom (see Methods). Figure 1d shows the white-light microscopy image of a typical device. The suspended 2D $MnPS_3$ membrane forms a drumhead resonator with a fundamental flexural-mode resonance frequency determined by the geometry, Young's modulus, and strain (see Supplementary Section S4). An RF voltage combined with a DC gate voltage is applied to the local back gate to electrically drive the membrane to vibrate, and the membrane oscillatory displacement can be detected through laser interferometry (Figure 1c). The measurement system details can be found in Methods. In this report, we focus on the mechanical fundamental resonance mode measurements. Figure 1e shows the resonance of a 25 nm-thick $MnPS_3$ drumhead, showing a very high $Q$ of ~56,000 at 4 K (device #1). When applying a DC back gate voltage, we can electrically tune the material's strain with the correspondingly exerted static Coulomb force. A DC gate voltage sweeping up to ±60 V yielded a $W$-shaped frequency tuning curve, which demonstrates the interplay between capacitive softening and elastic stiffening (Figure 1f). Combined with the extracted Young's modulus (see Supplementary Section S8, S13 and Methods) of this material, we estimate a lattice parameter change resolution down to the attometer ($10^{-18}m$) level, which corresponds to $\frac{\delta a}{a} \sim 10^{-9}$ ($a$: lattice constant) and membrane transverse displacement sensitivity of femtometer scale with such a high $Q$ factor.

## 2.2 Magnetic Field Dependence of Linear and Nonlinear Resonator Responses

An out-of-plane magnetic field was applied while measuring the mechanical resonance. We first examined the magnetic field dependence with the resonator operating in its linear regime with a perturbative small RF driving voltage (10 mV). **Figure 2a** shows the resonance frequency shift while sweeping the magnetic fields. In the range of -3T and +3T, the resonator frequency remained roughly constant. With further increasing field, the frequency showed an overall decrease. Most notably, two sharp frequency drops were observed at ±4.6 T ($H_1$) and at ±7 T ($H_2$), which are direct indications of sudden tension reduction and mark the presence of two distinct magnetic transitions. In comparison, when the device was warmed up above the Neel temperature (100 K, see Supplementary Figure S13), no sharp dip- or shoulder-like transitions were observed, confirming the magnetic origin of these two dips. The $H_1$ transition is close to the SF transition that was previously observed in neutron scattering and magnetic susceptibility measurements[31, 33, 35]. The spin configuration is expected to transit from AFM to canted states across the SF transition, as depicted in the Figure 2a insets. On the other hand, the $H_2$ transition has not been identified in direct magnetization characterization. While the SF transition is expected to be first-order in nature and possesses a hysteresis field, the theoretically calculated hysteresis gap is less than 0.02 T and thus does not contribute to the $H_2$ transition[36]. A previous magnetoresistance measurement on exfoliated $MnPS_3$ also reported a slope change in the tunneling magnetoresistance at around 7 T, which was at that time assigned to be the polarizing field from AFM to FM state[33]. Given the large exchange energy ($\mu_0 H_E \sim 100$ T)[32] of this material and our modeling of magnetostriction effects in the latter part of this article, the observed $H_2$ transition here should come from a different origin.

The two transitions show distinctly different hysteretic responses. The $H_1$ transition exhibits a sharper dip with a decreasing magnetic field sweeping, while the increasing field sweeping gives more shoulder-like gradual transitions. The $H_2$ transition, on the other hand, has no hysteresis and possesses a broader transition width compared to $H_1$. The extracted transition fields for devices of several different $MnPS_3$ thicknesses are plotted in Supplementary Figure S14 and summarized in Supplementary Table S1. Our results indicate that there is a slight decrease in the transition fields with decreasing thicknesses, which is consistent with prior reports of layer-dependence in $MnPS_3$[33] and indicates a decrease in magnetic interactions with the reducing layer number.

Accompanied by the frequency dips, the $Q$ factor decreases up to one order of magnitude at these two transition fields. The inverse of the $Q$ factor ($Q^{-1}$) reflects the damping in the mechanical resonance mode, and our observations indicate the opening-up of strong dissipation channels at $H_1$ and $H_2$. At the measurement temperature (4 K), dephasing is not expected to contribute significantly to the Lorentzian spectral width[37], and the $Q$ factor is directly related to the ring-down time $\tau$ of the resonator by $Q = \pi f_0 \tau$, where $f_0$ is the peak frequency. We can then estimate the time constant $\tau_M$ of the new dissipation channel by considering a $Q$ factor drop from $6 \times 10^4$ to $6 \times 10^3$ at $f_0 \sim 39$ MHz, which gives $\tau_M \sim 50$ μs at the $H_2$ transition.

We further examine the results when driving the MnPS$_3$ resonator at high RF power levels that induce Duffing nonlinear resonator behavior[38]. The restoring force of a mechanical resonator under a small displacement $z$ is written as $F(z) = F_0 + kz + k_3 z^3 + k_5 z^5 + \cdots$. When the higher-order terms of $k_3$ (and in fewer cases, of $k_5$) are included under a strong driving condition, the resonance peak becomes asymmetric with bifurcation hysteretic jumps between the forward and backward frequency sweeps, as depicted in **Figure 3a**. Figure. 3b displays the evolution of linear to nonlinear Duffing response as a function of the increasing RF drive voltage. Here, we use the frequency gap $\Delta f_{\text{hys}}$ between the hysteretic jumps to characterize the magnetic field dependence of Duffing nonlinearity. Figure 3d plots the $\Delta f_{\text{hys}}$ of another MnPS$_3$ NEMS resonator (device #2) with RF driving voltage of 400 mV. The two sharp transitions of $H_1$ and $H_2$ are clearly manifested in the nonlinear response. The prominent hysteretic behavior at $H_1$ is also consistent with the hysteresis in linear resonator frequency shifts in Figure 2a. As shown in Figure 3c, at the $H_1$ of 4.2 T, the resonance peaks completely lost the hysteretic frequency jumps compared to other fields outside the transition points. Such modification in nonlinearity is particularly abrupt for the downward field sweep across $H_1$. The nonlinear spring coefficients are analyzed in detail in Supplementary Section S9 using models ignoring the damping terms. The corresponding results are summarized in **Figure 4** f and g, which will be discussed later.

To understand the nature of the $H_1$ and $H_2$, we first estimate the expected magnetostriction effects based on the SF transition and previously measured magnetic interactions in MnPS$_3$. We model the AFM spin structure and the spin-flop transition as coupled two-spin sublattices with an easy axis along the out-of-plane direction, which is simplified from the slightly tilted easy axis extracted from the neutron scattering measurements[34]. The total magnetic energy $U_M$ before and after the SF transition field $H_{sf}$ can be written as

$$U_M = \begin{cases} -M_0 H_E, & H_\perp < H_{sf} \\ M_0(H_E \cdot \cos 2\theta - 2H_\perp \cdot \cos\theta + H_A \cdot \sin\theta^2), & H_\perp \geq H_{sf} \end{cases}$$

where $M_0$ is the sublattice magnetization, $H_E$ is the exchange field between the two spin sublattices[36]. $\theta$ is the spin canting angle as illustrated in Figure 4a, $H_A$ is the magnetic anisotropy field, and $H_\perp$ is the applied external magnetic field. In the ground state spin configuration, we have the spin canting angle $\theta$ to be $\cos\theta = \frac{H_\perp}{2H_E - H_A}$. The total energy of this spin-NEMS coupled system contains the elastic energy, work done by the boundary, and the magnetic energy. We can identify the strain change, and therefore the lattice parameter modifications, that minimizes the total energy as a function of the magnetic field. The full derivation is shown in the Methods section. Below the $H_{sf}$, $U_M$ is constant, yielding a constant strain $\varepsilon_0$ and minimal magnetoelastic effect. Above the $H_{sf}$, the strain $\varepsilon(H)$ will gain a quadratic magnetic field dependence of $\frac{2H_\perp^2}{3Y(2H_E - H_A)^2} \frac{\partial H_E}{\partial \varepsilon}$. Comparing the estimated magnetoelastic effects with our measured results, the frequency plateau between -3 T to +3 T and the rapid overall decrease with increasing fields can be attributed to the SF transition. The decrease in frequency, and therefore a decrease in strain, above the $H_{sf}$ infers that $\frac{\partial H_E}{\partial \varepsilon} < 0$. This is consistent with the expectation that spin exchange interactions decrease with a larger lattice separation.

The two sharp dips in the resonator linear and nonlinear responses are thus beyond the prediction of the simple SF model. To identify their origins, we now focus on the spin configuration near the SF transition field. Previous neutron scattering experiments extracted a large exchange energy of $\mu_0 H_E \sim 100$ T and a small anisotropy energy of $\mu_0 H_E \sim 0.06$ T [31-32]. After reaching the $H_{sf}$, the spin canting angle is calculated to have $\cos\theta = \frac{H_\perp}{2H_E - H_A}$. $\theta$ is therefore close to $\pm\frac{\pi}{2}$ and remains at this limit within our measured field range. As a result, above $H_{sf}$, the collinear AFM spin rotates to effectively a collinear in-plane spin alignment, with a large Neel component lying in the *ab* plane. However, minimal in-plane magnetic anisotropy was identified or predicted in the previous neutron scattering measurements[31-32] and theoretical studies[39]. The in-plane Neel vector component therefore does not have a preferred orientation. This leads to degenerate in-plane spin configuration, and we expect the formation of AFM domains above the SF transition field, as depicted in Figure 4a.

## 2.3 Theoretical Simulations

To verify this, the DW dynamics are evaluated by spin-dynamics simulations at different fields. The Supplementary Section S15 contains the description of simulation methods and results. Figure 4d-e shows the snapshot of the simulated Neel vector orientations near the spin-flop field ($H_{sf}$), which reveals the rich dynamic of the spin textures within nanoscale dimensions. Under an external field of 1.1 $H_{sf}$ (Figure 4e), simulations show multiple DWs merging and rapid DW movements, in comparison with the slower DW dynamics at other fields like 0.8 $H_{sf}$ (Figure 4d). The DW velocity can be estimated from the simulations, which peaks at $H_{sf}$ as shown in Supplementary Figure S21. Around the SF transition, substantial amounts of spin waves are emitted throughout the media which propagate with velocities of several km s$^{-1}$ (Supplementary Figure S21). The rapid DW motions show a strong layer dependence in the thin-film limit as a result of interlayer coupling. To estimate our measured devices' DW dynamics, we examine the extrapolated trend in Supplementary Figure S20, which yields > 600 m/s DW velocities for the measured device thicknesses. The corresponding frequency for DW lateral motion across the laser-detected area is then hundreds of MHz or higher, indicating multiple DW moving across the resonator within one mechanical vibration cycle. It is also expected from the simulations (see Supplementary Movies) that there are rapid domain emergence and annihilation within nanosecond timescales at the SF transition, which further increases the domain-induced dynamic frequency.

We attribute the sharp $H_1$ and $H_2$ transitions to the rapid and collective magnetic DW motions. Magnetic domains introduce additional magnetoelastic effects, which can be estimated with an additional magnetic domain wall energy $U_{DM}$ in the total NEMS system energy. With the presence of one DW, $U_{DW} = Js^2 \frac{\pi}{\sigma}$ where $\sigma$ is the domain wall width, and the exchange energy $J = H_E \mu_B$ [13]. At a transition field with rapid magnetic domain fluctuations, we model the system response as having a large number $N(t)$ of identical DWs, each carrying the energy of $U_{DM}$. The magnetostriction of the fluctuating DW number $N(t)$ can be evaluated by considering the time-averaged $\langle N \rangle$ within one mechanical vibration cycle, which yields a strain change of $\varepsilon(DW) - \varepsilon = \frac{1}{3Y}\frac{\partial}{\partial\varepsilon}[\langle N \rangle \cdot J]$. Since $\frac{\partial H_E}{\partial\varepsilon} < 0$, the presence of DW will cause a reduction in strain and a decrease in resonator frequency. The sharp frequency dips in Figure 2a thus originate from the strongly fluctuating DW dynamics. We note here that these types of AFM domains are conventionally difficult to detect due to the vanishing net magnetization. While it is possible to have magnetization at the DW, it is very challenging to perform nanoscale magnetometry with magnetic fields large enough to induce these transitions ( > 4 T) here.

The $H_1$ transition that is close to the expected $H_{sf}$ can be understood as the emergence and collective quenching of DWs and spin textures. On the other hand, the $H_2$ transition corresponds to the realignment of the in-plane Neel vector and the magnetic domains. The magnetic hysteresis in both the linear and nonlinear resonator measurements provides strong evidence to support the above assignment and further elucidate the nature of the two transitions. In the linear response regime, as shown in Figure 2a, the gradual decrease in frequency with an increasing field across $H_{sf}$ corresponds to the emergence of multiple domains. In contrast, the sharp frequency dip and, therefore, a narrow transition width when sweeping the field down past $H_{sf}$, infers a high $\langle N \rangle$ DW change and a collective quenching of magnetic domains. Remarkably, such hysteric behavior is only present if the applied field is swept past the $H_2$ transition, which is most evident from the nonlinear resonator measurements shown in Figure 4b and 4c (see also Supplementary Section S6 for the extraction from linear resonator measurements). The origin of domain realignment at $H_2$ can be qualitatively understood when we consider the AFM Neel vector's tilting angle of 8° at the zero field[34]. Under an out-of-plane magnetic field, the external field then has an effective projection perpendicular to the magnetic anisotropy direction, which eventually can assist in aligning the in-plane Neel vector direction along the tilting direction at higher fields. It is worth mentioning that there is still no conclusive microscopic model to understand the tilting of the easy-axis and the magnetic anisotropy in MnPS3. Several papers discussed the important roles of long-range interactions, like dipole-dipole interactions, as the possible origin of the anisotropy[40-42]. It remains an open question to fully understand the magnetic anisotropy in MnPS3 theoretically and subsequentially to understand the realignment of domains at 7 T.

## 2.4 Enhanced Mechanical Nonlinearity Coupled to Domain Dynamics

Finally, the abnormal Duffing nonlinearity at $H_1$ further reveal the impacts of complex spin texture. Under a high RF driving force, most of the mechanical responses, as plotted in Figure 3b, shows a stiffness hardening that is fitted with a cubic term $k_3 > 0$. Surprisingly, the resonance backbone curve at $H_1$ (Fig. 4h) exhibits a composite nonlinear behavior from softening to hardening, and the fitting requires the fifth-order term $k_5 z^5$ in the spring restoring force. The fitted $k_3$ has an abrupt sign flip, along with a significant enhancement in $k_5$, as summarized in Figure 4f and 4g at $H_1$. The sign flip and negative $k_3$ (spring softening) at $H_1$ is intriguing, because material-related nonlinearity mechanisms, like geometric nonlinearity and nonlinear damping[38], lead to spring hardening instead of a softening. The abnormal nonlinearity here is correlated with the rapid and complex spatial spin textures. The fluid-like DW motion from simulations near the SF transition potentially leads to nonlinear elastic mechanics due to the spatially varying and time-fluctuating magnetostriction forces. The strong nonlinearity from the enhanced fifth-order term $k_5 z^5$ infers a more chaotic spin-induced interaction and potentially fluid dynamics from the spin texture. At a sufficiently high drive level, the spring restoring force is dominated by the $k_5 z^5$ high-order term (Figure 4h), which suggests an increase in long-range spin interactions. The back action from layer rippling and slipping[43] under a high RF drive and, therefore, large displacement can further modify the interlayer spin configuration and nonlinear dynamics. An integrated microscopic model of the spin-induced mechanical nonlinearity and corresponding nonlinear dynamics will need further theoretical investigation.

## 3. Conclusion

In summary, we have identified hidden magnetic transitions in 2D AFM MnPS3 due to the collective dynamics of magnetic domain through nanomechanical resonator measurements. The abnormal mechanical nonlinearity modulation near the spin-flop field suggests nonlinear elastic and fluid-like spin texture behavior. The method established herein can be implemented to explore other vdW or thin-film systems hosting topologically non-trivial spin textures such as skyrmions, vortices, etc.. Our work paves the way towards mechanical manipulation of spin texture and harnessing the nonlinear coupling between the spin and lattice. The findings will inspire the next-generation nanomechanical devices coupled to spin textures for both classical and quantum information technologies[6-7].

## 4. Methods

## Sample Fabrication

We fabricate 4 drumhead resonators using few-layer $MnPS_3$ membranes by employing all-dry transfer techniques to avoid any contamination on the flakes. 23 nm Pt on 2 nm Ti is deposited and patterned onto a bare sapphire substrate to form local gates. 500 nm $SiO_2$ and 20 nm $Al_2O_3$ are deposited as dielectric layer using plasma enhanced chemical vapor deposition and atomic layer deposition techniques. Microtrenches with 520 nm depth and varying diameters are formed using reactive ion etching. 30nm Au on 5 nm Ti are deposited for source and drain electrodes and 200 nm Au on 5 nm Ti are deposited for contact pads. Finally, exfoliated $MnPS_3$ flakes are transferred to make the NEMS resonators (for additional information, see Supplementary Section S1). Atomic force microscopy is used to estimate the thickness of the transferred flakes (for details, see Supplementary Section S1).

## NEMS measurement setup

To precisely measure the resonance characteristics of the $MnPS_3$ membranes, we engineer and optimize a custom-built ultrasensitive optical interferometry system with a low-power 633 nm laser for readout of device motion (Fig. 1c). Such interferometry system is robust and can probe mechanical resonances both in linear and nonlinear regimes. A DC voltage from a source meter (Keithley 2400) is combined with an RF voltage from a network analyzer (HP3577A) using a bias-tee, and is applied to the local gate to drive the resonator. The interferometric signal contains information of the fm- to pm-scale vibrations of the membranes. A photodetector converts the optical signal to electronic signal that is read out by the network analyzer. More details are provided in Supplementary Section S3. Linear resonance measurement is performed at low RF voltage (~10 mV) and the nonlinear resonance characterization is performed at 400 mV. Initial device characterization is done at room temperature and in moderate vacuum (~20 mTorr). Magnetic field dependency is carried out at cryogenic temperature (~4 K) with the Attodry 1000 system (at the University of Florida) and also partially done at the National High Magnetic Field Laboratory in Tallahassee, Florida.

## Resonance Frequency Tuning and Estimation of Young's Modulus

Resonance frequency tuning of the fundamental mode of a drumhead resonator with initial strain $\varepsilon_0$, Poisson's ratio $\nu$, depth of the air gap $h_0$, and effective mass $m_{\text{eff}} = 0.2695\pi r^2 h\rho$ can be described by[44]

$$f_0(V_g) = \frac{1}{2\pi}\sqrt{\frac{\frac{\pi\epsilon_0^2 r^2}{8(1-\nu^2)E_Y h\varepsilon_0^2 h_0^4}(V_g-V_0)^4 + 4.9 E_Y h\varepsilon_0 - \frac{\epsilon_0\pi r^2}{3h_0^3}(V_g-V_0)^2}{m_{\text{eff}}} + \frac{\beta_0^4 D}{\rho h r^4}}$$

Here, for the first mode, $\beta_0^2 = 10.215$. Because of charge trapping in the suspended flake, the charge neutrality point shifts during the gate voltage sweeps and $V_0$ accounts for this effect. Young's modulus ($E_Y$) and the volume mass density ($\rho$) of the membrane can be found by fitting the experimentally measured resonance with varying gate voltage. The extracted values for resonator 1 are $E_Y$=770GPa, $\varepsilon_0$=0.0008, and $\rho$=2.5$\rho_0$. The volume mass density is 2.5 times higher than that of the expected value ($\rho_0$=2920 kg/m$^3$) likely due to the presence of surface adsorbates on the membrane at cryogenic temperatures. The extracted values for resonator 2 are $E_Y$=1659 GPa, $\varepsilon_0$=0.0007, and $\rho$=5.5$\rho_0$ and resonator 4 are $E_Y$=485 GPa, $\varepsilon_0$=0.0018, and $\rho$=3.8$\rho_0$. More details about fitting can be found in Supplementary Section S8.

## Magnetostriction in spin-flop transition

The total energy of the magnetic membrane system $U$ contains contributions from the magnetic terms $U_M$, elastic energy $U_{el}$ and the work done by the boundary $U_b$, $U = U_M + U_{el} + U_b$. We have $U_{el} = \frac{3}{2}E_Y\varepsilon^2$, where $E_Y$ is the effective Youn's modulus and $\varepsilon$ is the strain, and $U_b = -2U_{el}$ [29]. To estimate the magnetostriction of an SF transition, we consider the magnetic energy of AFM to canting spin states (Figure 4a inset) with a canting angle of $\theta$. In the AFM phase, $U_M = -H_E M_0$, where $H_E$ is exchange field and $M_0$ is the magnetization of a sublattice spin. In the spin-flop (canted) phase with a magnetic field of $H$ parallel to the easy axis, $U_M = H_E M_0 \cos 2\theta - 2HM_0 \cos\theta + H_A M_0 \sin^2\theta$. Since $\cos\theta = \frac{H}{2H_E - H_A}$, the magnetic energy can be further simplified to $U_M = \left(-H_E + H_A - \frac{H^2}{2H_E - H_A}\right) M_0$. The SF field can be solved: $H_{sf1} = \sqrt{H_A(2H_E - H_A)}$ and $H_{sf2} = \sqrt{H_A(2H_E + H_A)}$. Since $H_A \ll H_E$, their differences can be ignored, and the magnetic energy change is continuous from the AFM to the SF state.

The strain $\varepsilon$ as a function of the magnetic field can be found out by minimizing the total energy $\frac{\partial U}{\partial \varepsilon} = 0$, which yields $-3E_Y\varepsilon + \frac{\partial U_M}{\partial \varepsilon} = 0$. Since $U_M$ does not depend on the magnetic field in the AFM state, the strain is a constant. In the SF state, the strain change compared to the AFM state is then $\frac{3E_Y}{M_0}(\varepsilon_{AFM} - \varepsilon_{SF}) = -\frac{\partial H_A}{\partial \varepsilon}\left(1 - \frac{H^2}{(2H_E - H_A)^2}\right) - \frac{\partial H_E}{\partial \varepsilon}\frac{2H^2}{(2H_E - H_A)^2} \approx = -\frac{\partial H_A}{\partial \varepsilon} - \frac{\partial H_E}{\partial \varepsilon}\frac{2H^2}{(2H_E - H_A)^2}$.

**Estimation of responsivity and fitting of nonlinear resonance signals**

The critical amplitude ($a_c$) for onset of nonlinearity of a drumhead resonator is given by $a_c = \sqrt{\frac{8\sqrt{3}}{9k_3^2 Q}}$, where the Duffing coefficient can be estimated by $k_3^2 = \frac{13 + 21\nu - 4\nu^2}{30(1+\nu)r^2\varepsilon_0}$. We estimate $a_c = 0.91$ nm at 4 K and 0 T. From the nonlinearity measurement, we also estimate the signal caused by the critical amplitude $v_c = 58$ μV in the voltage domain. From these two, we obtain the displacement-to-voltage responsivity (transduction gain) of the fundamental mode to be 64 nV/pm. The backbone curve of the nonlinear response can be expressed as[45]

$$\Omega = \omega_0 + \frac{3k_3 z_0^2}{8\omega_0 m_{\text{eff}}} + \frac{5k_5 z_0^4}{16\omega_0 m_{\text{eff}}}$$

where $k_3$ and $k_5$ denote the Duffing and quintic nonlinear coefficients, $\Omega$, $m_{\text{eff}}$, and $z_0$ represent the instantaneous resonance frequency, effective mass, and amplitude, respectively. Using this equation, we can estimate $k_3$ and $k_5$ as fitting parameters from the calibrated displacement measurement. We can also estimate $k_3$ and $k_5$ directly in the voltage domain without the displacement calibration with units N/V$^3$ and N/V$^5$, respectively. The trend of $k_3$ and $k_5$ matches qualitatively in both voltage and displacement domains.

**Atomistic spin dynamics**

Spin dynamics simulations methods[46-50] were used to calculate the magnetic properties of MnPS$_3$. The system was described using the spin Hamiltonian:

$$\mathcal{H} = -\frac{1}{2}\sum_{i \neq j} J_{ij} \mathbf{S}_i \cdot \mathbf{S}_j - \frac{1}{2} B \sum_{NN; i \neq j} (\mathbf{S}_i \cdot \mathbf{S}_j)^2 - K \sum_i (S_i^z)^2$$

where $\mathbf{S}_{i,j}$ are unit vectors representing the local spin directions on sites $i,j$. The first term represents bilinear exchange with the exchange constant $J_{ij}$ between spins $i$ and $j$. The bilinear exchange was considered up to the third nearest neighbor for in-plane and out-of-plane neighbors. The second term describes the biquadratic exchange contribution between sites $i$ and its in-plane nearest neighbors and $B$ is the biquadratic exchange constant. Both the bilinear and biquadratic interactions are isotropic. The last term is the easy-axis anisotropy contribution with the magnetocrystalline constant $K$. Supplementary Table 2 gives the exchange parameters used for the calculations. Additional details are included in Supplementary Section S15.

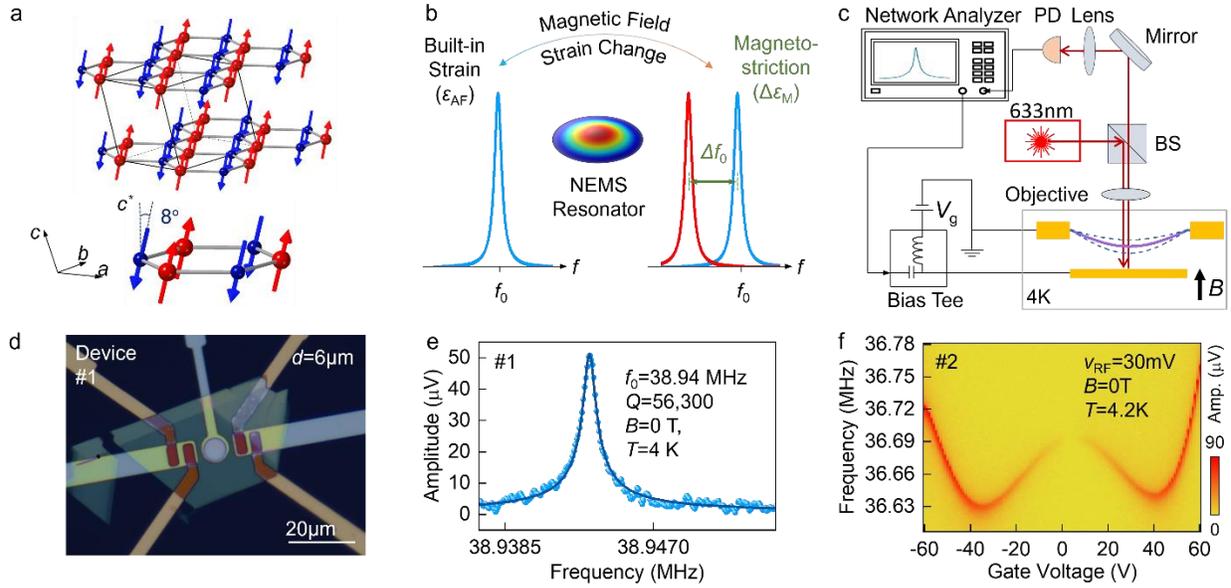

**Figure 1| MnPS$_3$ nanoelectromechanical resonator devices. a,** The crystal structure of few-layer MnPS$_3$. The spin orientation of Mn atoms is represented by the up and down arrows. **b,** Conceptual illustration of magnetostriction induced resonance frequency shift due to spin-flop transition in MnPS$_3$ NEMS resonators. Due to magnetostriction effect, the strain in the membrane changes leading to a downshift in the resonance frequency of the NEMS resonator. **c,** Optical interferometry system to measure the driven response of the resonators. BS: beam splitter, PD: photodetector. The 4 K cryostat has a superconducting magnet that can sweep the magnetic field in both directions. The applied magnetic field is in out of plane direction of the suspended membrane. **d,** Optical image of MnPS$_3$ nanoelectromechanical resonator. Scale bar: 20 μm. **e,** Measured resonance in MnPS$_3$ NEMS resonator (device #1) with $d$=6 μm. Fitting to a finite-$Q$ damped harmonic resonator model exhibits resonance at $f_0$=38.94 MHz with $Q$=56,300 at 4 K. **f,** Resonance frequency tuning by varying the DC gate voltage (device #2). Resonance frequency tuning demonstrates a $W$-shaped tuning curve with wider $|V_G|$ sweep range. The number with # in any panel denotes the device identification number used in this study.

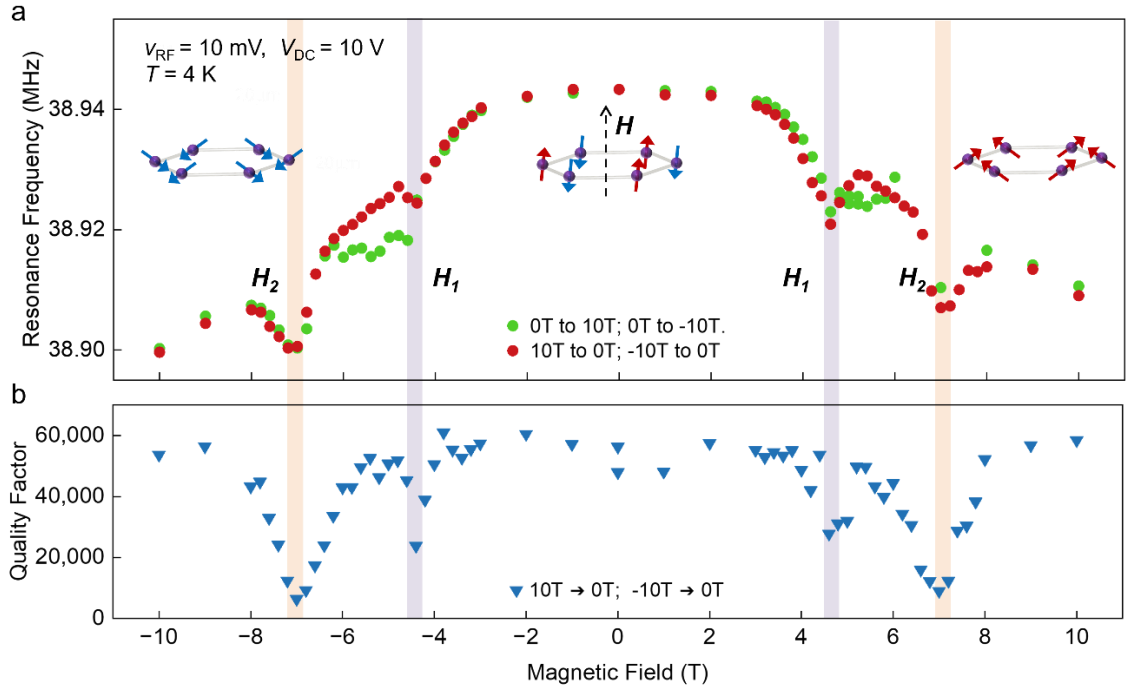

**Figure 2| Magnetic field dependence of mechanical resonator in linear region. a,** Evolution of resonance frequency as a function of out-of-plane magnetic field measured between ±10 T. A complete cycle is achieved by sequentially adjusting the magnetic field $\mu_0 H$ from 0 T to 10 T, then decreased from 10 T to -10 T, and finally returned to 0 T. Sharp transitions are observed near $\mu_0 H_1 = \pm 4.6$ T and $\mu_0 H_2 = \pm 7$ T, as indicated by the purple and orange stripes. The insets plot the corresponding spin configurations: AFM below the SF transition (middle) and canted-spin states above the SF transition (left and right). **b,** Measured $Q$ factor versus magnetic field shows strong dissipation at the sharp transitions.

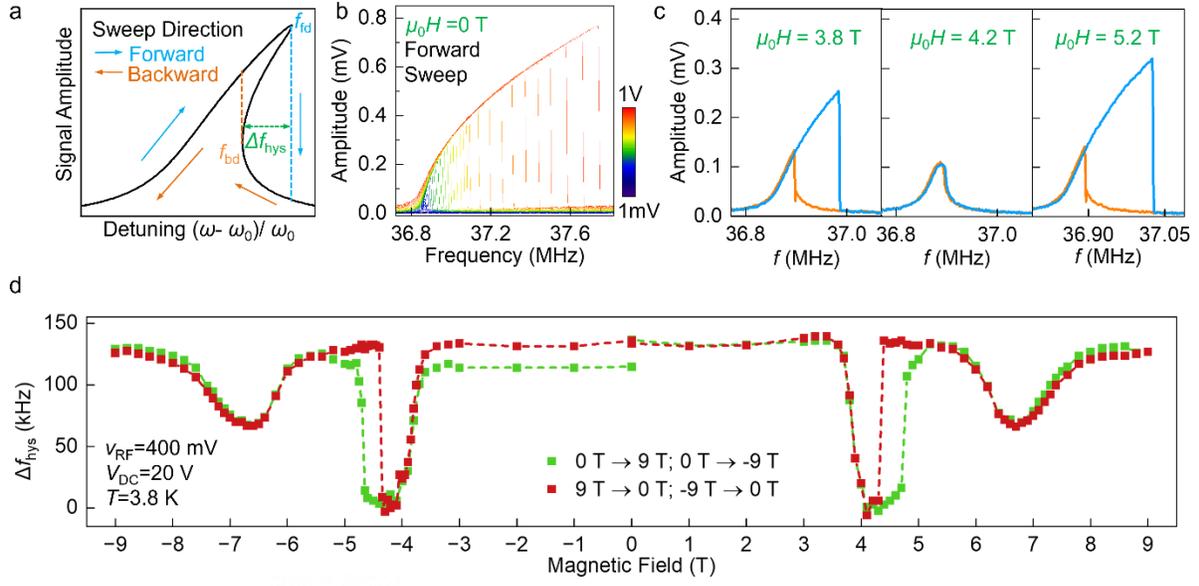

**Figure. 3| Magnetic field dependence of Duffing nonlinearity. a,** Conceptual illustration of Duffing nonlinearity with increased drive force. Forward and backward sweeps show a frequency hysteresis due to Saddle Node bifurcation. **b,** Duffing nonlinearity measurement by increasing $v_{RF}$ from 1 mV to 1 V at zero field. **c,** Evolution of Duffing nonlinearity in device 2 at varying magnetic field. At 4.2 T, the forward and backward sweeps coincide with each other. **d,** Measured frequency hysteresis ($\Delta f_{hys}$) during forward and backward frequency sweeps versus out-of-plane magnetic field shows characteristic sharp dips at the transition fields. The transition at 4.2 T exhibits hysteresis during increasing and decreasing magnetic fields.

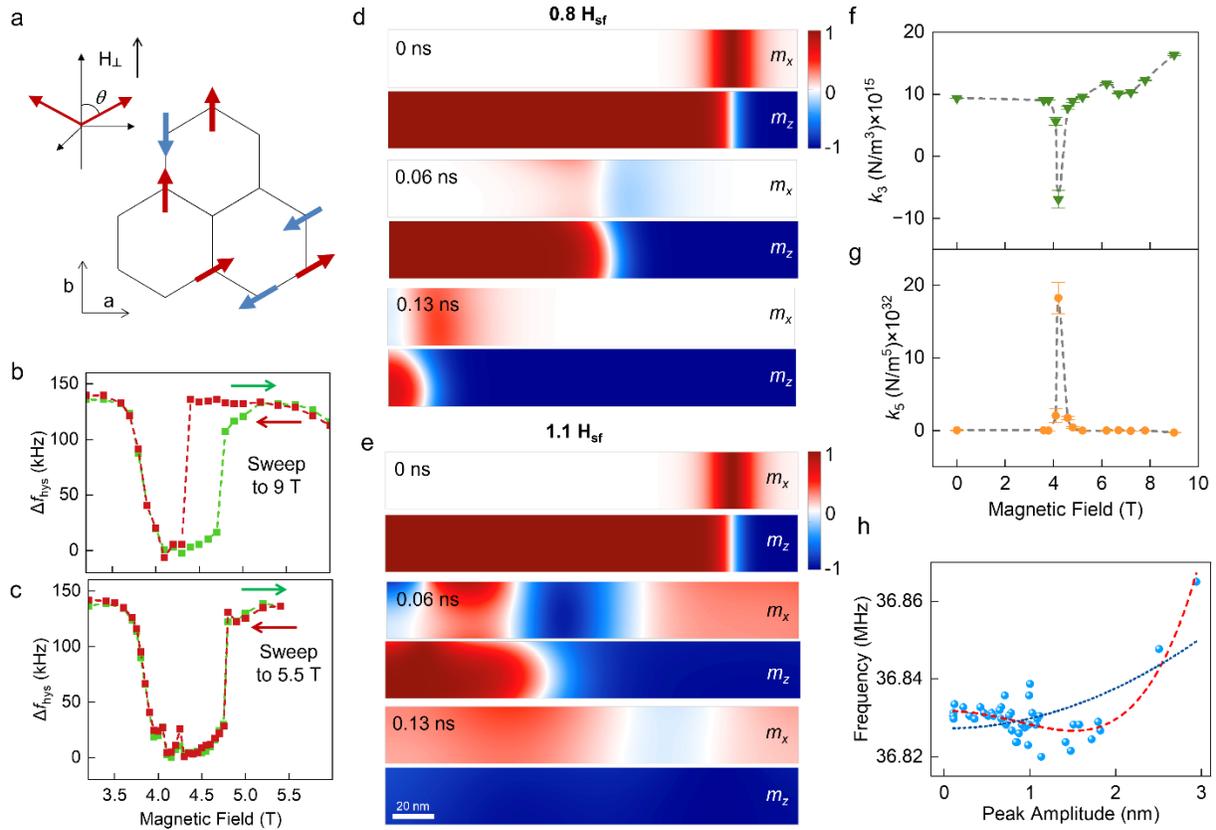

**Figure 4| Complex magnetic domains in magnetic transition. a,** Possible in-plane spin orientations above the spin-flop field. Due to large exchange and small anisotropy, the canting angle in a spin-flop state, as plotted in the upper right corner, is close to 90º under a moderate out-of-plane magnetic field. **b** and **c** compare the different hysteresis of nonlinear signals. The hysteresis is missing when the field sweep is below $H_2$. **d-e,** Spin dynamics simulations for AFM domains at magnetic fields of 0.8 $H_{sf}$ and 1.1 $H_{sf}$, respectively, of the spin-flop field ($H_{sf}$) projected along the $m_x$ and $m_z$ components of the AFM magnetization at different time-steps (0 ns, 0.06 ns, 0.13 ns). A domain-wall is initially stabilized at the system with an odd number of layers (5L) and its dynamics is followed at 4.2 K. **f-g,** The magnetic field dependence of $k_3$ (in **f**) and $k_5$ (in **g**) from the backbone curve. **h,** The backbone curve at $\mu_0 H_1 = 4.2$ T. The red dashed line is the fitting that includes $k_5$ and the blue dashed line is the fitting that includes only up to $k_3$. The resonator frequency curve at $H_1$ thus change from spring hardening to a softening to hardening trend.

## Data availability

The data that support the findings of this study are available within the paper and its Supplementary Information. The input files for the VAMPIRE simulations for the various layers and fields are available at the repository: https://github.com/shreyas-ramachandran/mnps3_domain_wall_motion. Additional data are available from the corresponding authors upon request.

## Acknowledgments

X.-X. Z. acknowledges the support from the Department of Energy (DOE) award DE-SC0022983, which provides the experimental instrument and supplies expenses. X.-X. Z. and P. F. are thankful to the National Science Foundation (NSF) through the QuSeC-TAQS Program (Grant OSI-2326528). P. F. also thanks the NSF for support through the IUSE Program (Grant DUE-2142552). A portion of this work was performed at the National High Magnetic Field Laboratory, which is supported by National Science Foundation Cooperative Agreement No. DMR-2128556* and the State of Florida. S.R. acknowledges Ricardo Rama-Eiroa for valuable comments on the simulations. E.J.G.S. acknowledges computational resources through CIRRUS Tier-2 HPC Service (ec131 Cirrus Project) at EPCC (http://www.cirrus.ac.uk), which is funded by the University of Edinburgh and EPSRC (EP/P020267/1); and ARCHER2 UK National Supercomputing Service via the UKCP consortium (Project e89) funded by EPSRC grant ref EP/X035891/1. E.J.G.S. acknowledges the EPSRC Open Fellowship (EP/T021578/1), the Donostia International Physics Centre for funding support.

## Author contribution

X.-X.Z. and P.F. conceived the experiments. S.E.Y. and Y.W. fabricated the devices and performed data analysis. S.E.Y., Y.W., and J.K-P. performed the experiments. L.S., S.M., C.T. and D.S. assisted the measurements done in National High Magnetic Field Lab. E.J.G.S. designed the theoretical approach. S.R. performed spin dynamics simulations under E.J.G.S. supervision. X.-X.Z. and S.E.Y. wrote the manuscript with critical inputs from P.F., E.J.G.S., S.R. All authors discussed the results and commented on the manuscript.

## Competing Interests

The authors declare no competing financial interest.